\documentclass{emulateapj}
\singlespace
\slugcomment{Accepted for the publication in the {\it Astrophysical Journal} (Scheduled on Sep 10, 2007, v666)}
\def\farcm{\hbox{$.\mkern-4mu^\prime$}}

\def\la{\mathrel{\hbox{\rlap{\hbox{\lower4pt\hbox{$\sim$}}}\hbox{$<$}}}}
\def\ga{\mathrel{\hbox{\rlap{\hbox{\lower4pt\hbox{$\sim$}}}\hbox{$>$}}}}

\shortauthors{Park}
\shorttitle{G299.2$-$2.9}

\begin{document}

\title{{\it Chandra} X-Ray Study of Galactic Supernova Remnant G299.2$-$2.9}

\author{Sangwook Park}

\affil{Department of Astronomy and Astrophysics,
525 Davey Lab., Pennsylvania State University, University Park, PA. 16802} 
\email{park@astro.psu.edu}

\author{Patrick O. Slane}
\affil{Harvard-Smithsonian Center for Astrophysics, 60 Garden Street,
Cambridge, MA. 02138}

\author{John P. Hughes}
\affil{Department of Physics and Astronomy, Rutgers University,
136 Frelinghuysen Road, Piscataway, NJ. 08854-8019}

\author{Koji Mori}
\affil{Department of Applied Physics, University of Miyazaki, 1-1 Gakuen 
Kibana-dai Nishi, Miyazaki, 889-2192, Japan}

\and

\author{David N. Burrows and Gordon P. Garmire}
\affil{Department of Astronomy and Astrophysics,
525 Davey Lab., Pennsylvania State University, University Park, PA. 16802}

\begin{abstract}

We report on observations of the Galactic supernova remnant (SNR) G299.2$-$2.9 
with the {\it Chandra X-Ray Observatory}. The high resolution images with 
{\it Chandra} resolve the X-ray-bright knots, shell, and diffuse emission
extending beyond the bright shell. Interior to the X-ray shell is faint 
diffuse emission occupying the central regions of the SNR. Spatially-resolved 
spectroscopy indicates a large foreground absorption ($N_{\rm H}$ $\sim$ 
3.5 $\times$ 10$^{21}$ cm$^{-2}$), which supports a relatively distant 
location ($d$ $\sim$ 5 kpc) for the SNR. The blast wave is encountering a highly 
inhomogeneous ambient medium with the densities ranging over more than an order 
of magnitude ($n_0$ $\sim$ 0.1 $-$ 4 cm$^{-3}$). Assuming the distance of 
$d$ $\sim$ 5 kpc, we derive a Sedov age of $\tau$ $\sim$ 4500 yr and an
explosion energy of $E_0$ $\sim$ 1.6 $\times$ 10$^{50}$ ergs. The ambient 
density structure and the overall morphology suggest that G299.2$-$2.9 
may be a limb-brightened partial shell extending to $\sim$7 pc radius 
surrounded by fainter emission extending beyond that to a radius of $\sim$9 pc. 
This suggests the SNR exploded in a region of space where there is a density
gradient whose direction lies roughly along the line of sight. The faint central 
region shows strong line emission from heavy elements of Si and Fe, which is 
caused by the presence of the overabundant stellar ejecta there. We find no 
evidence for stellar ejecta enriched in light elements of O and Ne. The observed 
abundance structure of the metal-rich ejecta supports a Type Ia origin for 
G299.2$-$2.9.

\end{abstract}


\section {\label {sec:intro} INTRODUCTION}

High angular resolution imaging spectroscopy with the {\it Chandra X-Ray 
Observatory} has revolutionized the study of supernova remnants (SNRs)
(e.g., Weisskopf \& Hughes 2005 for a recent review); spatially-resolved 
spectral analysis with arcsecond-scale resolution imaging provides a powerful 
tool to reveal a number of new and detailed structures in the SNRs. 
We continue such an X-ray study of SNRs with the {\it Chandra} observation 
of SNR G299.2$-$2.9.

The Galactic SNR G299.2$-$2.9 is one of the handful number of SNRs that were
discovered in X-rays with the {\it ROSAT} all sky survey \citep{buss95}.
The X-ray morphology reported with the previous X-ray detectors on board
{\it Einstein}, {\it ROSAT}, and {\it ASCA} appeared to be 
``centrally-enhanced'' with a partial shell-like structure 
\cite{buss95,slane96,bai00}. The X-ray spectrum was dominated by thermal 
emission without evidence for nonthermal synchrotron emission from
the embedded pulsar wind nebula (PWN) at the SNR center \cite{bai00}. 
A thermal composite SNR was thus suggested for G299.2$-$2.9, with 
centrally-enhanced X-ray emission produced by thermal evaporation of 
cloudlets \citep{white91}. Although there were some controversial 
interpretations of the SNR dynamics (young and nearby vs. middle-aged and 
distant) with the early {\it ROSAT} PSPC data \citep{buss96}, subsequent 
works found that a model assuming a middle-age ($\tau$ $\sim$ 6600 yr) SNR 
at a distance of $d$ $\sim$ 5 kpc self-consistently produces the observed 
X-ray and IR fluxes \cite{slane96}. The results from spectral analysis with 
{\it ASCA} data also supported the interpretation of a Sedov phase SNR at 
$d$ $\sim$ 5 kpc \citep{bai00}. 

With the relatively small angular size of the SNR ($\sim$15$'$ diameter, Busser 
et al. 1996) and the poor angular resolution of the data used by the previous 
authors, the true morphology of G299.2$-$2.9 could not be resolved. 
The {\it ASCA} data allowed only the analysis of the integrated spectrum from 
the overall SNR. We note that unpublished archival {\it ROSAT} HRI image 
shows that G299.2$-$2.9 is in fact shell-type with bright enhancements in 
the northeastern portion of the shell (\S~3), which most likely 
caused the morphological confusion by previous authors (shell-type vs. 
centrally-enhanced). However, no spectral information is available with 
the HRI data. Spatially-resolved spectroscopy with high resolution imaging 
is essential to unveil the true nature of the SNR. Here we present the results 
from the {\it Chandra} observation of SNR G299.2$-$2.9. We describe our 
{\it Chandra} observation and the data reduction in \S~2. 
Results from the imaging and the spectral analyses are presented in 
\S~3 and \S~4, respectively. We discuss 
the interpretations in \S~5, and summarize the conclusions in 
\S~6.

\section{\label{sec:obs} OBSERVATION \& DATA REDUCTION}

We observed G299.2$-$2.9 with the Advanced CCD Imaging Spectrometer
(ACIS) \citep{garmire03} on board {\it Chandra} on 2005 April 9 as part
of the {\it Chandra} Guaranteed Time Observation program. In order to 
focus on the detailed study of the complex, bright, soft X-ray 
structure in the northeastern parts of the SNR, we used the ACIS-S3
as the primary detector by pointing at the brightest X-ray knot in
the northeast. A large portion of the SNR was thus placed on the S3 chip
to provide the highest available response at low energies, where the
characterization of emission from O-K and Fe-L lines is particularly
important for the SNR studies. Considering the angular size of 
$\sim$15$^{\prime}$ for G299.2$-$2.9, we also turned on other CCDs 
to cover the SNR as much as possible beyond the S3 chip boundary.
As we show in Fig.~\ref{fig:fig1}, portions of the SNR extend onto
adjacent CCDs, providing nearly full coverage of the SNR. No severe 
variability was found in the background light curve. We corrected the 
spatial and spectral degradation of the ACIS data caused by radiation 
damage, known as the charge transfer inefficiency (CTI; Townsley et al. 
2000), with the methods developed by Townsley et al. (2002a), before 
further standard data screening by status, grade, and energy selections. 
``Flaring'' pixels were removed and {\it ASCA} grades (02346) were selected. 
After the data reduction, the effective exposure is 29.6 ks. The overall 
SNR spectrum is soft and there are few source photons above $E$ $\sim$ 
3 keV. At very low energies ($E$ $\la$ 0.4 keV), the source flux is 
negligible because of the foreground absorption, and X-ray emission is 
mostly dominated by the detector background. Thus, photons between 
0.4 and 3.0 keV are extracted for the data analysis. 

\section{\label{sec:image} X-RAY MORPHOLOGY }

\begin{figure}[]
\figurenum{1}
\centerline{\includegraphics[angle=0,width=0.9\textwidth]{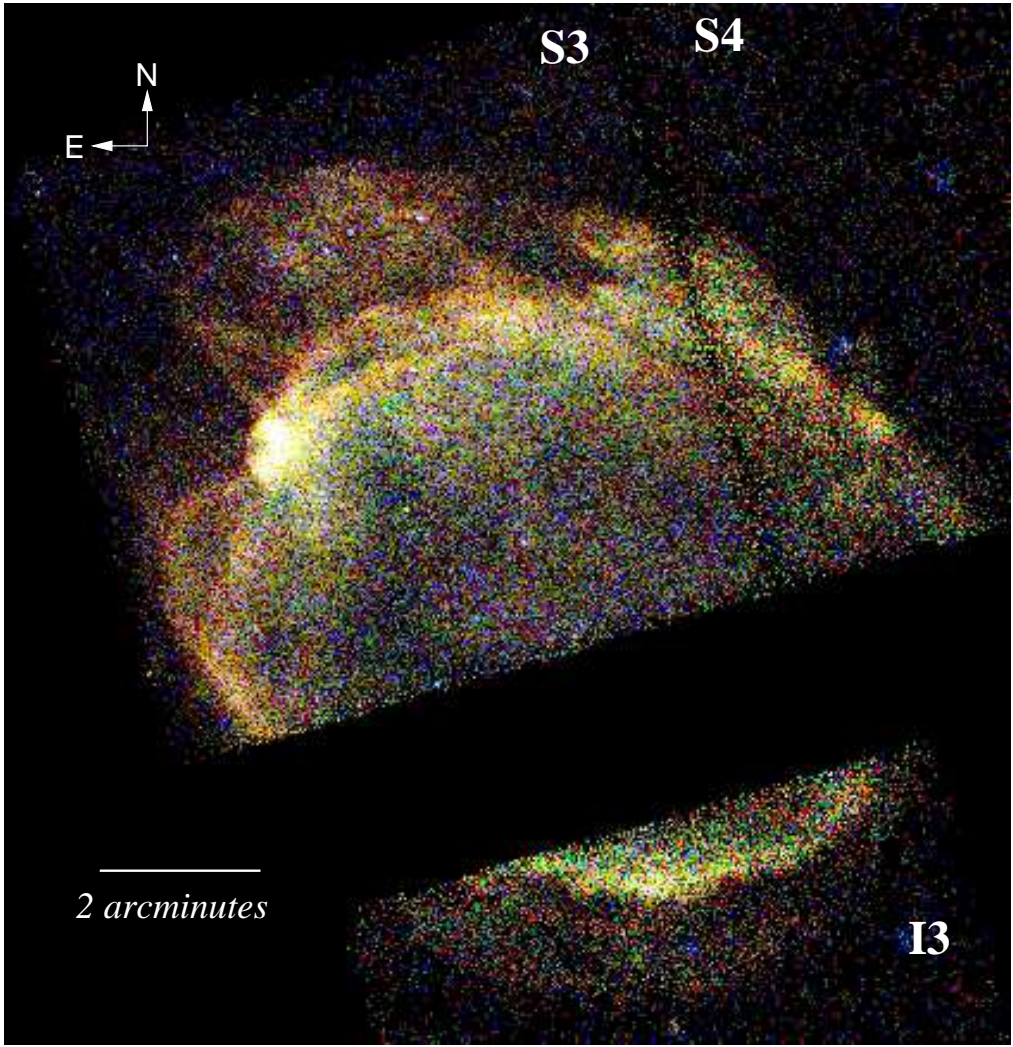}}
\figcaption[]{Three-color image of G299.2$-$2.9: Red is 0.4 $-$ 0.9 keV,
green is 0.9 $-$ 1.4 keV, and blue is 1.4 $-$ 3.0 keV. Each subband image 
has been corrected for the exposure and the detector efficiency. 
The images have been binned by 4 pixels for the purpose of display. 
\label{fig:fig1}}
\end{figure}

The ``X-ray-color'' {\it Chandra} ACIS image of G299.2$-$2.9 is presented 
in Fig.~\ref{fig:fig1}. Our primary detector is the ACIS-S3 with the 
pointing roughly at the brightest X-ray knot in the northeastern part 
of the X-ray shell, and thus the SNR is for the most part detected on 
the ACIS-S3. Portions of the southern and the western boundaries are also 
detected on the ACIS-I3 and S4 chips. The I3 data are particularly useful 
by clearly imaging the bright southern shell and diffuse emission 
extending beyond the shell, which is similar to the features in the 
northern side of the SNR. 

\begin{figure}[]
\figurenum{2}
\centerline{\includegraphics[angle=0,width=0.7\textwidth]{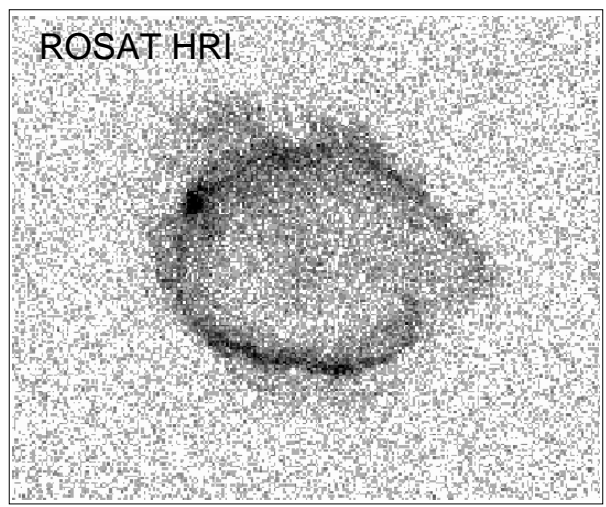}}
\figcaption[]{The archival {\it ROSAT} HRI image of
G299.2$-$2.9 (taken from the {\it SkyView} service by the HEASARC), 
showing the SNR in its entirety.
\label{fig:fig2}}
\end{figure}

\begin{figure}[]
\figurenum{3}
\centerline{\includegraphics[angle=0,width=0.8\textwidth]{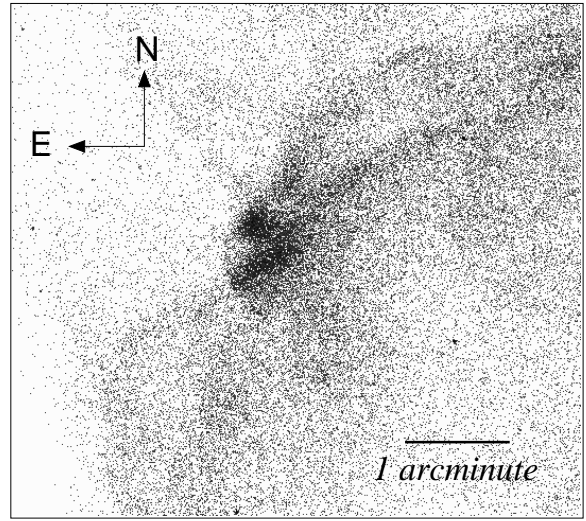}}
\figcaption[]{The full-resolution (0$\farcs$5) broadband ACIS image of the
region around the brightest knots in the northeastern shell. Darker grey-scales
correspond to higher intensities.
\label{fig:fig3}}
\end{figure}

Fig.~\ref{fig:fig1} shows that the overall morphology of the SNR is shell-type. 
We note that unpublished archival {\it ROSAT} HRI data also reveal the 
shell-type X-ray morphology of G299.2$-$2.9 (Fig.~\ref{fig:fig2}). The 
brightness distribution of the {\it ROSAT} HRI image is dominated by a 
bright, almost complete shell, beyond which more diffuse emission extends 
to fairly large distances, in particular in the northeast. In the center of 
the image a low brightness patch of diffuse emission stands up above the 
general SNR background emission. Our {\it Chandra} image shows that the X-ray 
shell is generally soft (red to yellow in Fig.~\ref{fig:fig1}), and reveals
complex structures such as multiple and/or broken shells, particularly 
in the eastern and northern sides. In general, these relatively bright, 
soft shell features appear to make a nearly complete circumferential shell 
around the SNR with an angular extent of $\sim$10$^{\prime}$ in the 
east-west and $\sim$9$\farcm$3 in the north-south direction. The brightest 
X-ray feature in the northeastern shell of the SNR is resolved into two 
small knots ($\sim$20$^{\prime\prime}$$-$40$^{\prime\prime}$ angular sizes) 
and surrounding extended features ($\sim$1$'$ sizes) (Fig.~\ref{fig:fig3}). 
These knots are located at the ``connecting'' or ``overlapping'' position of 
multiple shells. This morphology suggests that these bright knots may 
represent dense interstellar medium (ISM) along the path of the blast 
wave, rather than metal-rich stellar ejecta. The forward shock appears
to be interacting with and wrapping around these dense, small ISM clouds.
The peculiar morphology of the two bright knots surrounded by shock fronts
probably suggests that the small clouds might have also been crushed by 
the interacting strong shock, as proposed for a similar structure
seen in Puppis A SNR \citep{hwang05}. 

Faint diffuse emission extends beyond the bright shell and appears to 
extend almost all around the SNR, particularly in the northern and southern 
portions of the SNR (Fig.~\ref{fig:fig1}). The angular distance between 
the northern and the southern boundaries of diffuse extended emission is 
$\sim$13$^{\prime}$. These faint emission features are also spectrally 
soft and, for the most part, appear to make the outer boundary of the SNR. 
We note that both the bright X-ray shell and faint diffuse emission 
extending beyond the shell appear to be nearly complete around the entire 
SNR. This overall morphology suggests that the blast wave is likely 
expanding into an inhomogeneous ISM, making two distinctive, large-scale 
forward shock fronts. 

\begin{figure}[]
\figurenum{4}
\centerline{\includegraphics[angle=0,width=0.7\textwidth]{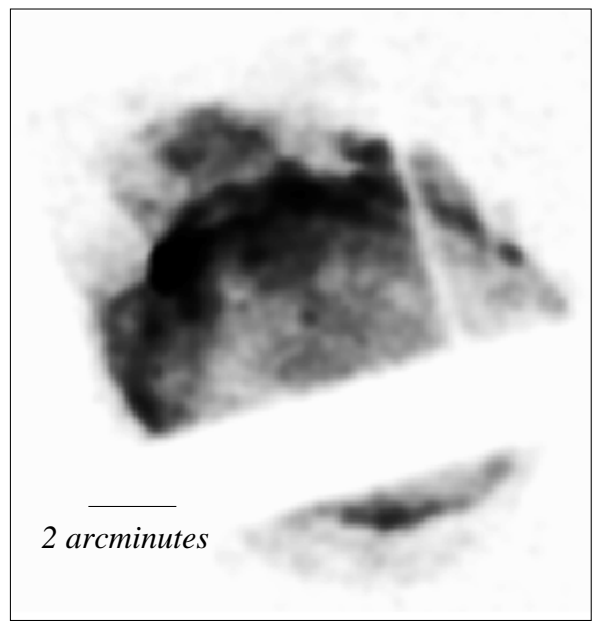}}
\figcaption[]{The combined image of Fe (0.75 $-$ 0.95 keV) and Si 
(1.75 $-$ 1.95 keV) lines of G299.2$-$2.9. The image has been binned into 
6 $\times$ 6 pixels and then smoothed with a Gaussian ($\sigma$ = 3 pixels) 
for the purposes of display. In order to emphasize the faint central ejecta 
nebula, the bright shell regions are saturated in dark grey-scales. 
\label{fig:fig4}}
\end{figure}

We detect faint diffuse emission in the central region of the SNR 
(Fig.~\ref{fig:fig1}). This central emission feature is enhanced in the 
Fe-L and Si-K lines (\S~4), which show a central nebulosity 
(Fig.~\ref{fig:fig4}). The {\it Chandra} image suggests that the faint 
central emission feature has an angular size of $\sim$4$'$. Because of the 
low surface brightness and the partial coverage with the ACIS, however, it is
difficult to accurately estimate the angular size of this central feature. 
We note that this central feature is reminiscent of the metal-rich ejecta 
nebula discovered in some middle-aged SNRs in the Magellanic Clouds 
\citep{hughes03,park03a,park03b}. Indeed, our spectral analysis of the SNR 
indicates the presence of overabundant heavy elements in the central nebula 
of G299.2$-$2.9 (\S~4).

\section{\label{sec:spec} Spectral Analysis}

\begin{figure}[]
\figurenum{5}
\centerline{\includegraphics[angle=0,width=\textwidth]{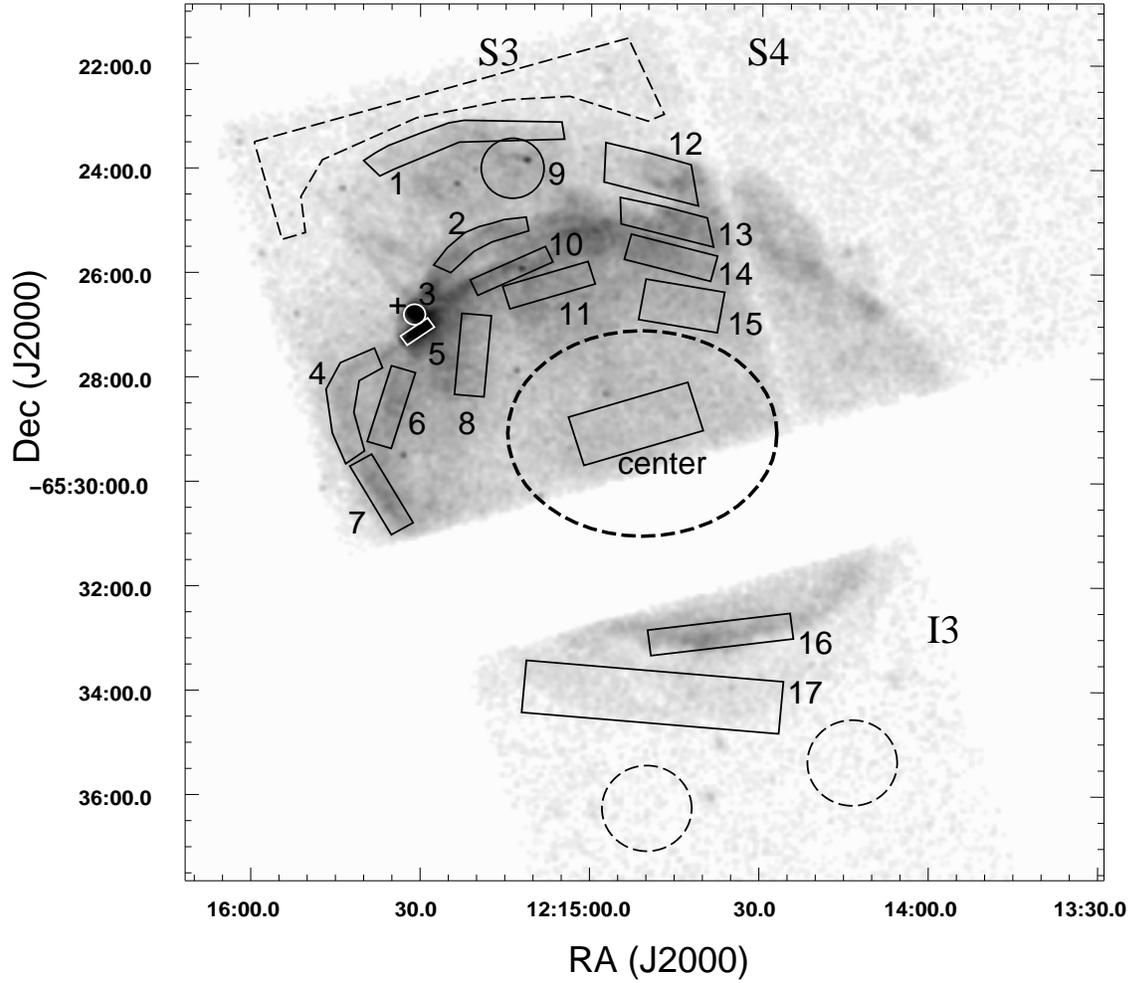}}
\figcaption[]{The gray-scale broad band (0.4 $-$ 3.0 keV) image
of G299.2$-$2.9 from the {\it Chandra}/ACIS observation. 
The image has been binned with 2 pixels and then
smoothed by a Gaussian ($\sigma$ = 5 pixels) for the purposes of display.
The small regions used for the spectral analysis (Table~\ref{tbl:tab1}) 
are marked with solid boxes and circles. 
The thick dashed ellipse schematically indicates the faint diffuse 
central emission feature enriched in Si and Fe. The pointing direction
is marked with a cross, just outside of the brightest knot (region 3)
on the S3 chip. The background was estimated with emission from 
the thin dashed regions. 
\label{fig:fig5}}
\end{figure}

Spectral studies with {\it Chandra} have revealed that X-ray emission from 
SNRs is rarely described by a simple spectrum such as a single-temperature 
thermal plasma or the Sedov model \citep[e.g.,][and references therein]{weiss05}. 
Instead, the X-ray emission from SNRs typically originates from 
multiple-temperature plasmas in various ionization stages caused by 
nonequilibrium conditions in blast wave shocks propagating into an ambient 
medium with a wide range of interstellar density. X-ray emission from shocked, 
metal-rich ejecta is often detected with modern X-ray telescopes, 
in middle-aged SNRs as well as in young SNRs. Spatially-resolved spectral 
analysis from small regions within an SNR, rather than analysis of the overall 
integrated spectrum, is thus essential for the proper study of SNRs. We extract 
a number of regional spectra from G299.2$-$2.9 utilizing our ACIS data. There 
are $\sim$30 faint detected pointlike sources in the ACIS field of view. 
These point sources have been removed before extracting regional spectra. 
Depending on regional X-ray intensities and morphologies, we selected regions 
with angular sizes from $\sim$20$^{\prime\prime}$ up to a few arcminutes. We 
choose our regions to provide $\sim$3000 or more counts in each spectrum in 
order to place reasonable constraints on the fitted spectral parameters. 
We mark the representative regions of our spectral analysis in Fig.~\ref{fig:fig5}. 

The background spectrum is estimated from source-free regions outside of 
the SNR's northern (for regions on the S3 chip) and southern (for regions 
on the I3 chip) boundaries. Each regional source spectrum has been binned 
to contain a minimum of 20 counts per channel. For the spectral analysis 
of our CTI-corrected data, we use the response matrices appropriate for the 
spectral redistribution of the CCD, as generated by Townsley et al. (2002b). 
The CIAO {\tt caldb version 3.21} is used for the data processing. We use a 
non-equilibrium ionization (NEI) plane-parallel shock model \citep{bor01} 
({\tt vpshock} in conjunction with the NEI version 2.0 in the {\tt XSPEC}) which 
is based on {\tt ATOMDB} \citep{smith01}. We use an updated version of this
atomic database to include inner-shell processes that are missing in the current 
{\tt XSPEC} NEI version 2.0\footnote{The unpublished version of the updated 
model has been provided by K. Borkowski.}. The contribution of line emission 
from elements He, C, N, Ca, Ar, and Ni is insignificant in the fitted energy 
range ($E$ = 0.4$-$3.0 keV); thus we fix the abundances of these elements at 
the solar values \citep{anders89} (hereafter, elemental abundances are with 
respect to the solar). Abundances of other species (O, Ne, Mg, Si, S, and Fe) 
are varied freely in the fits (S abundance was occasionally fixed at 1 in cases 
where photon statistics are low at $E$ $\sim$ 2$-$3 keV).

\begin{figure}[]
\figurenum{6}
\centerline{\includegraphics[angle=0,width=0.9\textwidth]{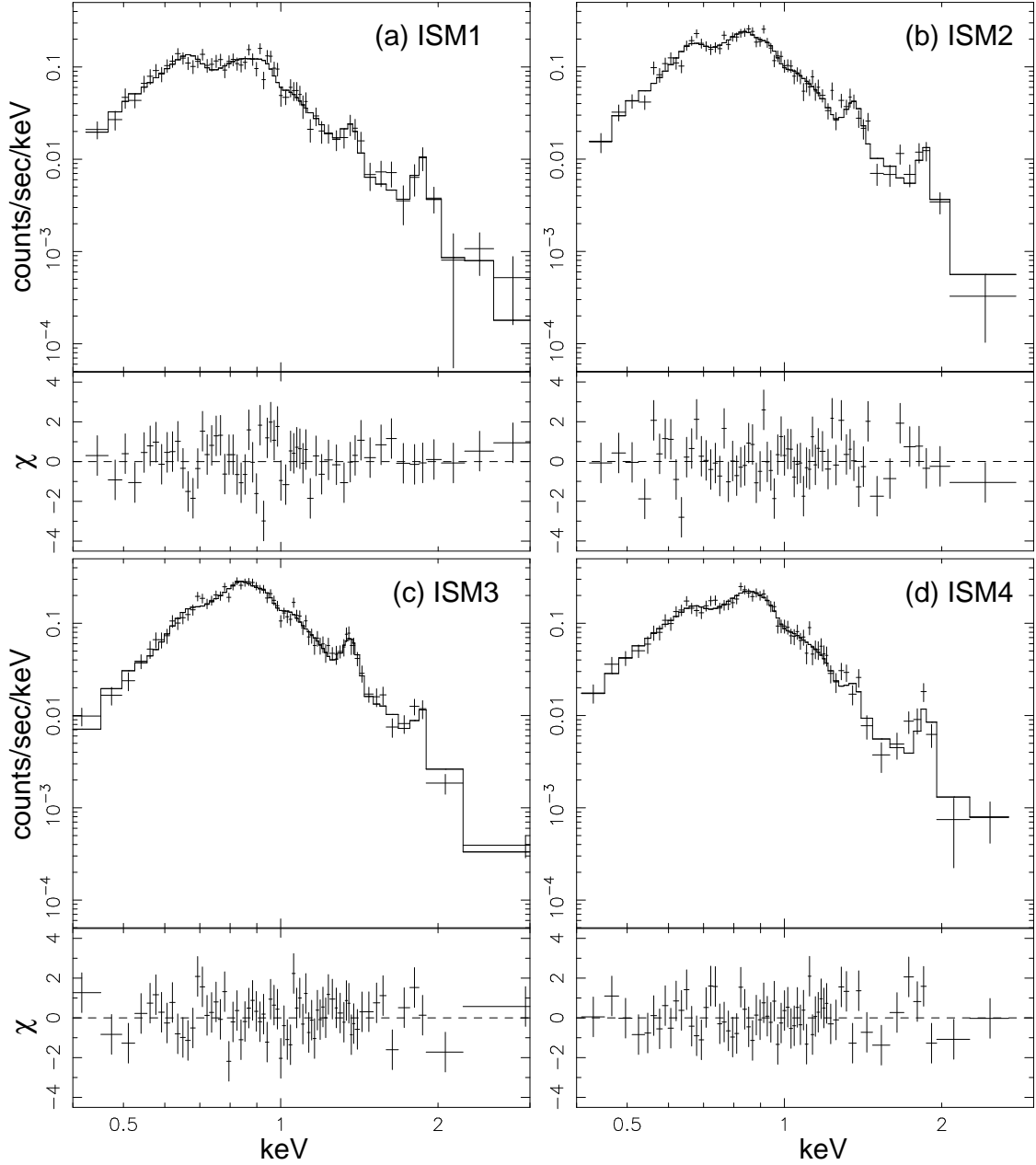}}
\figcaption[]{The regional ACIS spectra of G299.2$-$2.9. Four representative
spectra from the shocked ISM regions. Panels (a), (b), (c), and (d) 
correspond to regions 1, 2, 3, and 4 in Fig.~\ref{fig:fig5}, respectively. 
The best-fit planar shock model is overlaid in each panel. 
The lower plot in each panel is the residuals from the best-fit model.
\label{fig:fig6}}
\end{figure}

X-ray spectra from regions along the bright shell, and from diffuse emission beyond 
the shell, show characteristics consistent with emission from shock-heated ISM; 
on average, the thermal plasma has an electron temperature $kT$ $\sim$ 0.55 keV 
and ionization timescale $n_{\rm e}t$ $\sim$ 4.5 $\times$ 10$^{11}$ cm$^{-3}$ s 
(Fig.~\ref{fig:fig6} and Table~\ref{tbl:tab1}; errors are 90\% confidence level, 
hereafter, unless noted otherwise.). The best-fit metal abundances are 
$\sim$0.5$-$0.7 for individual regions, which are marginally constrained within 
a factor of $\sim$2. The average values of abundances from these regions are 
O = 0.63$\pm$0.28, Ne = 0.65$\pm$0.22, Mg = 0.54$\pm$0.19, Si = 0.67$\pm$0.27, 
and Fe = 0.66$\pm$0.28 (1$\sigma$ errors for average values).

The absorbing column density ($N_{\rm H}$) ranges from $\sim$2 to $\sim$5 
$\times$ 10$^{21}$ cm$^{-2}$ (Table~\ref{tbl:tab1}). The extremely low 
column of $N_{\rm H}$ $\sim$ 10$^{20}$ cm$^{-2}$ suggested by earlier {\it 
ROSAT} data cannot fit the {\it Chandra} data, and is firmly rejected. 
Because of large statistical uncertainties (e.g., the estimated errors in 
Table~\ref{tbl:tab1} are typically $\sim$20$-$30 \% after fixing elemental 
abundances at the best-fit values), it is not clear whether the $N_{\rm H}$ 
variation is the true extinction variation across the SNR or 
statistical/systematic artifacts of the spectral fitting. For simplicity, 
we consider the average $N_{\rm H}$ $\sim$ 3.5 $\times$ 10$^{21}$ cm$^{-2}$ 
in the following discussion. (We confirm that spectral fits with $N_{\rm H}$ 
fixed at this average value would not change the results with the quality 
of the current data.) 

\begin{figure}[]
\figurenum{7}
\centerline{\includegraphics[angle=0,width=0.8\textwidth]{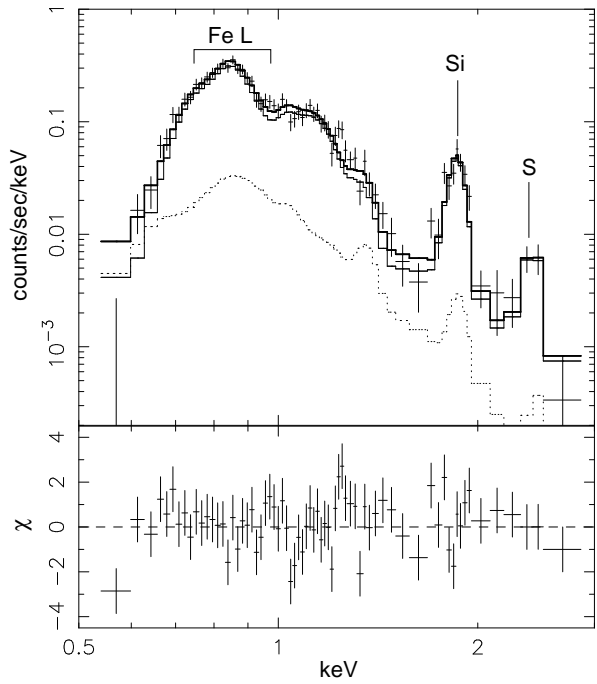}}
\figcaption[]{The spectrum of G299.2$-$2.9 from the center region. 
The dotted line is the model for the contribution from the projected
shocked ISM component, and the thin solid line is the central ejecta
component. The best-fit model (ejecta + ISM) is overlaid with the thick 
solid line. The lower plot is the residuals from the best-fit model.
\label{fig:fig7}}
\end{figure}

In contrast to the spectra from peripheral regions, faint emission from 
the central region reveals strong line features primarily from K-shell 
transitions in highly-ionized He-like Si ($E$ $\sim$ 1.85 keV) and 
L-shell transitions in low-ionization-state Fe ($E$ $\sim$ 0.8 $-$ 1 keV) 
(Fig.~\ref{fig:fig7}). The S-K line also appears to be enhanced. The spectral 
analysis shows strongly overabundant Si, S, and Fe for the central region, 
which indicates the presence of metal-rich ejecta material produced in the 
core of the progenitor star. In the spectral fitting of the center region, 
we added an additional plane shock component in order to account for the 
projected contribution from the swept-up ISM in the observed spectrum. 
Spectral parameters for the underlying ISM component are fixed at the average 
values derived from our regional spectral analysis of the circumferential area 
(i.e., $kT$ = 0.55 keV, $n_{\rm e}t$ = 4.5 $\times$ 10$^{11}$ cm$^{-3}$ s, 
O = 0.63, Ne = 0.65, Mg = 0.54, Si = 0.67, Fe = 0.66). The $N_{\rm H}$ was 
fitted, but tied in common between the ejecta and the ISM components. 
The best-fit $N_{\rm H}$ $\sim$ 3.7 $\times$ 10$^{21}$ cm$^{-2}$ is in good 
agreement with the average value obtained from other regional spectral analysis.
Only the normalization of the ISM component is freely varied in the fit. 
Results from the spectral fit of the central region are presented in 
Table~\ref{tbl:tab2}. We note that the fit is not particularly good from a 
statistical point of view ($\chi^2_{\nu}$ $\sim$ 1.4). This is not surprising 
for fits to complex spectra from SNRs, especially those from ejecta-dominated 
emission. There are uncertainties in the currently available atomic data 
as well as in detector calibration due to gain variation, particularly near 
the absorption edges and in the soft energy band ($E$ $\la$ 1 keV). In fact,
the relatively high $\chi^2$ value is dominated by a single lowest energy
bin ($E$ $\sim$ 0.5 keV) where the CCD calibration is less reliable. Detailed 
weak emission lines from the Fe-L complex are usually difficult to fit with 
low-resolution CCD spectra. The relatively large $\chi^2$ in the central 
region spectral fit is most likely caused by these factors, which is beyond 
the usual goodness-of-fit criterion. On the other hand, the main spectral 
features of the broad Fe-L line emission feature and the strong Si and S 
lines are well-fit and provide robust measurements of spectral parameters and 
the abundances of key elements (O, Fe, and Si). 

\section{\label{sec:disc} Discussion}

\subsection{\label{subsec:ism} Blast Wave and Interstellar Environment}

We have no reliable distance estimate to G299.2$-$2.9. There is no H{\small I} 
absorption distance measurement, which would be difficult anyway because of the 
faintness of the SNR in the radio band ($\sim$340 mJy at 4.85 GHz, Slane et al. 
1996). The SNR's foreground column density as measured from the spectral fit 
of our {\it Chandra} data, $N_{\rm H} \sim 3.5 \times 10^{21}$ cm$^{-2}$, is 
about half of the total Galactic $N_{\rm H}$ in this direction \citep{dl90}. 
This suggests that the SNR is neither very nearby ($d < 1$ kpc) nor extremely 
distant ($d > 10$ kpc). The line-of-sight in the direction of G299.2$-$2.9 
passes through the Carina-Sagittarius spiral arm twice: once at a distance of 
$\sim$2 kpc and then later at a distance of $\sim$11 kpc. Assuming that the
SNR is in the spiral arm or in the interarm region would be broadly consistent 
with the measured $N_{\rm H}$ distance constraint. Based on this limited amount 
of available independent data, we assign a plausible distance range of $d$ 
$\sim$ 2$-$11 kpc to G299.2$-$2.9. For concreteness, we scale numerical values 
to $d_5$ $\equiv$ $d\over{5~{\rm kpc}}$ in the following discussion, while it 
should be noted that the distance is uncertain by probably a factor of $\sim$2.

\begin{figure}[]
\figurenum{8}
\centerline{\includegraphics[angle=0,width=0.5\textwidth]{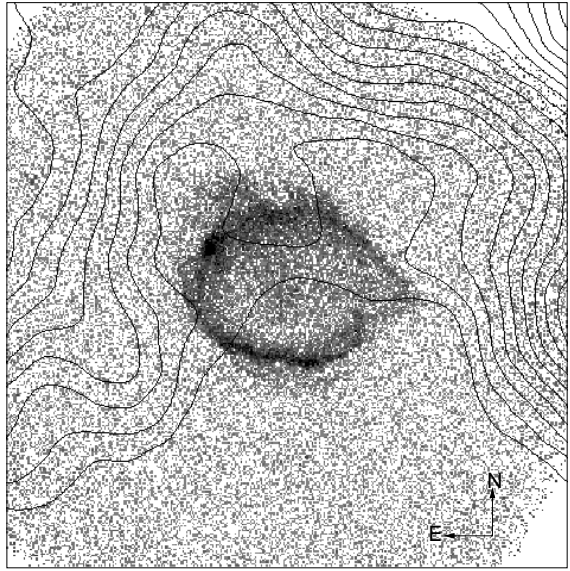}}
\figcaption[]{The gray-scale {\it ROSAT} HRI image of G299.2$-$2.9, overlaid with 
the {\it IRAS} 100 $\mu$m contours. The {\it IRAS} 100 $\mu$m 
image was taken from the {\it SkyView} services provided by the HEASARC. 
The Galactic plane is toward the north.
\label{fig:fig8}}
\end{figure}

\begin{figure}[]
\figurenum{9}
\centerline{\includegraphics[angle=0,width=0.5\textwidth]{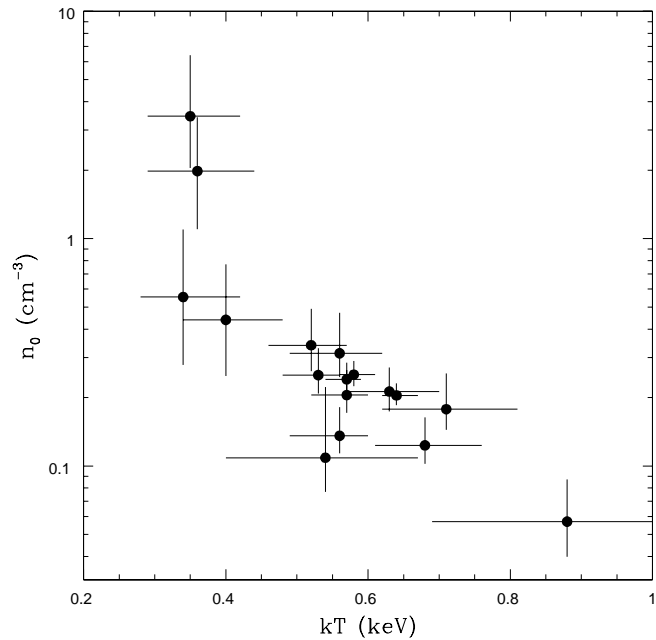}}
\figcaption[]{$kT$ vs. $n_0$ distribution
from the best-fit spectral models of the regional spectra of G299.2$-$2.9,
taken from the regions marked in Fig.~\ref{fig:fig5}. $n_0$ is calculated from 
the best-fit $EM$ from each regional spectral fit as described in the text.
17 regions (out of the total 21) showing normal abundances are presented.
The errors are estimated with the elemental abundances fixed at the best-fit values.
\label{fig:fig9}}
\end{figure}

X-ray spectra from around the shell show characteristics of hot thermal 
plasma from the shocked ambient ISM with normal composition. Various 
morphologies (e.g., the shell, filaments, knots and diffuse extended emission 
features with various surface brightness) then indicate that the blast wave 
is encountering highly inhomogeneous ambient densities. In fact, the {\it IRAS} 
100 $\mu$m map shows a complex interstellar environment around G299.2$-$2.9 
(Fig.~\ref{fig:fig8}). The interaction of the shock with a variable density 
structure may also have caused the observed variation in the electron temperature 
and the ionization state across the SNR. The preshock density vs. electron 
temperature distribution from the regional spectra is largely anti-correlated 
(Fig.~\ref{fig:fig9}). In fact, the correlation coefficient for all 17 data 
points is $-$0.61, and is $-$0.82 if the three extreme cases (regions 3, 5, 
and 17) are excluded. These values of the correlation coefficient imply a 
significant anti-correlation between the density and electron temperature 
(at $>$99\% C.L.). The preshock density $n_{\rm 0}$ from each region is derived 
from the best-fit volume emission measure $EM$ (= $n_{\rm e}$$n_{\rm H}V$, 
where $n_{\rm e}$, $n_{\rm H}$, and $V$ are the postshock electron density, 
the H density, and the emission volume assuming $d$ = 5 kpc, respectively). 
In this estimate, we assumed $n_{\rm e}$ = 1.2$n_{\rm H}$ for a mean charge 
state with normal composition and $n_{\rm H}$ = 4$n_{\rm 0}$ for the strong 
shock. For the volume estimates, we assumed that the pathlengths through 
the regions along the line of sight are comparable with the physical sizes 
corresponding to the angular lengths of the regions at $d$ = 5 kpc.  
The brightest knots in the northeastern part of the shell represent low 
temperatures of $kT$ $\sim$ 0.35 keV with the shock front entering the 
densest ISM, $n_{\rm 0}$ $\sim$ (2$-$4$) f^{-{1\over2}}$ cm$^{-3}$, 
where $f$ is the volume filling factor of the X-ray emitting gas. Along the 
X-ray shell, except for the brightest northeastern knots, preshock densities 
of $n_{\rm 0}$ $\sim$ (0.2$-$0.5)$f^{-{1\over2}}$ cm$^{-3}$ are implied. 
Faint diffuse emission in the northern, eastern, and southern boundaries 
indicates a high-temperature shock ($kT$ $\sim$ 0.7 keV) propagating 
into a lower density ISM ($n_{\rm 0}$ $\la$ 0.1$f^{-{1\over2}}$ cm$^{-3}$). 
These results indicate that the blast wave of G299.2$-$2.9 is expanding into 
more than an order of magnitude density gradient, including the densest 
small clouds in the northeastern side. We note that our density estimates 
are dependent on uncertainties involved in volume estimates. However, the 
volume effect is small ($n_0$ $\propto$ $V^{-{1\over2}}$), and thus our 
conclusions are not strongly affected unless our volume estimates are 
off by more than an order of magnitude.

The thin, bright filamentary emission features appear to make up a nearly 
complete circumferential shell around the entire SNR. On the other hand, 
faint diffuse emission extending beyond the bright shell forms the 
outermost boundary of the SNR in the north/northeastern and southern sides. 
Faint diffuse emission may also form the outer boundary along the western 
side, as suggested by the {\it ROSAT} HRI image (this region of the SNR was 
not covered by our {\it Chandra} observation). These two large-scale 
structures with the implied high and low densities suggest that the supernova 
might have exploded near the boundary between two regions of different ISM 
densities (on average, $n_{\rm 0}$ $\la$ 0.1 cm$^{-3}$ in one side and 
$n_{\rm 0}$ $\ga$ 0.3 cm$^{-3}$ in the other side). Each ``hemisphere'' 
might then be expanding into the relatively higher and lower ambient densities 
with low and high velocities, respectively, producing the observed morphology 
when seen in superposition along the line of sight (Fig.~\ref{fig:fig10}). 
This particular morphological development has been demonstrated by hydrodynamic 
simulations \citep{tan85}. The Galactic SNR VRO 42.05.01 (G166.0+4.3) is 
believed to be a prototype of such a case with an edge-on orientation, which 
shows a peculiar two-partial shell morphology in the radio band \citep{pin87}. 
The observed X-ray morphology of G299.9$-$2.9 appears to be the face-on view 
of the radio morphology of VRO 42.05.01. We note that, unlike G299.2$-$2.9, 
the peculiar morphology of two half-shells of VRO 42.05.01 is present only 
in the radio band. In X-rays, VRO 42.05.01 shows an enhancement only near 
the center of the SNR (Burrows \& Guo 1994; Guo \& Burrows 1997; Mori et al. 
in preparation). This is probably because of the large difference in ages 
between G299.2$-$2.9 and VRO 42.05.01; the X-ray shells of the older SNR 
VRO 42.05.01 ($\tau$ $\sim$ 20000 yr) might have cooled and no longer 
produce significant X-rays, while the relatively young ($\tau$ $\sim$ 4500 yr, 
see below for the age estimate) G299.2$-$2.9's shock is still in its hot 
adiabatic phase. The available archival data of G299.2$-$2.9 from the radio 
observations have only limited angular resolution of a few arcmins (e.g., 
Slane et al. 1996), and thus it is difficult to confirm whether the radio 
morphology of G299.2$-$2.9 also shows the same structure as seen in X-rays. 
Follow-up high resolution radio imaging observations should be helpful to 
reveal the nature of G299.2$-$2.9. 

The angular sizes ($\sim$9$\farcm$9 for the bright shell, $\sim$12$\farcm$6 
for the diffuse boundary, on average) imply physical sizes of $\sim$14.4 
$d_5$ pc and $\sim$18.4 $d_5$ pc for the bright shell and the diffuse outer 
boundary, respectively. These large sizes of G299.2$-$2.9 suggest a Sedov 
phase for the SNR. Based on the Sedov solution, we calculate some SNR 
parameters of G299.2$-$2.9 for the two representative cases of the smaller 
bright shell and the larger faint rim. The derived SNR ages are $\tau$ 
$\sim$ 4700~$d_5$ yr and 4200~$d_5$ yr for the large faint rim and the 
small bright shell, respectively. These results are generally consistent 
between the two large-scale structures. The explosion energies are estimated 
to be $E_0$ $\sim$ 1.6~$\times$ 10$^{50}$ $d^{5\over2}_5$ ergs for the large 
faint rim and $E_0$ $\sim$ 1.7~$\times$ 10$^{50}$ $d^{5\over2}_5$ ergs for the 
small bright shell. These $E_0$ values are also consistent between the two 
cases. G299.2$-$2.9 is thus likely an $\sim$(4000$-$5000)~$d_5$ yr old 
middle-age SNR with an explosion energy of $E_0$ $\sim$ 1.6~$\times$ 10$^{50}$ 
$d^{5\over2}_5$ ergs. The total swept-up mass, $M_{\rm swept}$, is 
$\sim$11$f^{1\over2}$$d^{5\over2}_5$ $M_{\odot}$. The SNR parameters are 
summarized in Table~\ref{tbl:tab3}. 

\subsection{\label{subsec:ejecta} Metal-Rich Ejecta}

The central emission feature is faint, and appears to have an average radius of 
$\sim$2$^{\prime}$, showing spectral characteristics entirely different from
those of the circumferential regions. The electron temperature is high ($kT$ 
$\sim$ 1.36 keV) with a low ionization timescale ($n_{\rm e}t$ $\sim$ 1.3 
$\times$ 10$^{10}$ cm$^{-3}$ s). The strong line emission from the elemental 
species of Si, S, and Fe reveals enhancements in the abundances of those elements. 
These results indicate that the central region of G299.2$-$2.9 is most likely 
the ejecta nebula, i.e., emission from low density metal-rich ejecta material 
heated by the reverse shock. Our spectral analysis indicates that the ejecta 
is enriched particularly in the heavy Si-group species produced in the deep 
core of the progenitor star. The detection of Si- and Fe-rich ejecta in 
the central region of the SNR is reminiscent of the Type Ia SNR DEM L71 in 
the Large Magellanic Cloud \citep{hughes03}. In analogy to DEM L71, we 
propose that G299.2$-$2.9 is a Type Ia SNR. 

To investigate this suggestion further, we compare the observed Si to Fe mass 
ratio with the standard Type Ia/II supernova nucleosynthesis models, estimating 
the mass ratio $M_{\rm Si}$/$M_{\rm Fe}$ from the best-fit $EM$ ratio of the 
center region. Based on the dominant ionization states of the Si and Fe implied 
by the observed spectrum, we assume that the electron densities associated with 
the Si and Fe ions are $n_{\rm e,Si}$ $\sim$ 12~$n_{\rm Si}$ and $n_{\rm e,Fe}$ 
$\sim$ 18~$n_{\rm Fe}$, for simple ``pure'' Si and Fe ejecta cases, 
respectively.  For the dominant isotopes of $^{28}$Si and $^{56}$Fe, 
the measured Si and Fe abundances imply $M_{\rm Si}$/$M_{\rm Fe}$ = 
0.52$^{+0.20}_{-0.15}$($V_{\rm Si}/V_{\rm Fe}$)$^{1\over2}$. Therefore, 
if the Fe and Si occupy the same size volumes ($V_{\rm Fe}$ = $V_{\rm Si}$), 
the mass ratio is $M_{\rm Si}$/$M_{\rm Fe}$ $\sim$ 0.52. In the case of DEM L71,
$V_{\rm Si}$ is a thick shell extending only about half way to the center 
of the SNR, while Fe is present even at the center of the SNR. If we assume 
such a stratified ejecta structure for G299.2$-$2.9, $V_{\rm Si}/V_{\rm Fe}$ 
is $\sim$0.88 and thus $M_{\rm Si}$/$M_{\rm Fe}$ $\sim$ 0.49. If $V_{\rm Si}$ 
corresponds to a thinner shell, $M_{\rm Si}$/$M_{\rm Fe}$ becomes only smaller. 
These small Si to Fe mass ratios are in good agreement with Type Ia supernova 
models \citep{nomoto97b}. For Type II cases, the Si to Fe mass ratio is 
typically large ($\sim$1$-$10, for progenitor masses of 18$-$70 $M_{\odot}$) 
\citep{nomoto97a}. We note that, in the case of a {\it complete} spatial mixture 
of Si and Fe (i.e., $n_{\rm e,Si}$ $\approx$ $n_{\rm e,Fe}$), 
$M_{\rm Si}$/$M_{\rm Fe}$ becomes $\sim$20\% smaller. Such a case may be 
implied by the assumption of $V_{\rm Fe}$ = $V_{\rm Si}$, but the actual 
$n_{\rm e,Si}$ and $n_{\rm e,Fe}$ distributions may depend on the details 
of the geometrical structure of the ejecta material. Nonetheless, these 
details of the ejecta structure would not affect our general conclusions. 
We also note that the mixture of H in the metal-rich ejecta may not affect 
our mass ratio estimates, as long as the H mixture rates are comparable between 
the Fe and the Si ejecta.

Although $M_{\rm Si}$/$M_{\rm Fe}$ of $\la$0.5 may also be consistent with 
a Type II case for a relatively less massive progenitor ($\la$15 $M_{\odot}$) 
\citep{nomoto97a}, the lack of any ejecta enriched in the O-group elements
argues against a Type II origin. Unlike the case of the Si- and Fe-rich 
ejecta, we find no evidence of enhancements for the light species, O, Ne, 
and Mg. For instance, the fitted O abundance is negligible ($<$0.1), 
suggesting that the Fe to O abundance ratio is high ($>$ 4). The Si to O 
abundance ratio is also high ($>$ 3). These high Fe and Si abundances relative 
to the O abundance are consistent with Type Ia cases ($\sim$1$-$5), while 
Fe/O and Si/O ratios are at least an order of magnitude lower for Type II 
supernovae \citep{nomoto97a,nomoto97b}. 

\section{\label{sec:sum} Summary and Conclusions}

\begin{figure}[]
\figurenum{10}
\centerline{\includegraphics[angle=0,width=0.7\textwidth]{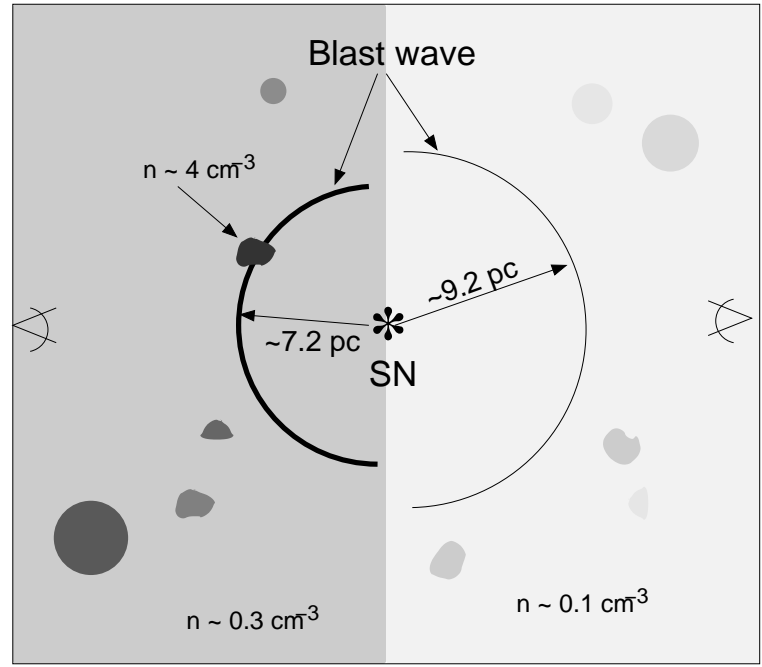}}
\figcaption[]{A schematic ``edge-on'' view of G299.2$-$2.9.
\label{fig:fig10}}
\end{figure}

We present results from the {\it Chandra} observation of the  Galactic
SNR G299.2$-$2.9. High resolution images from the {\it Chandra} observation 
confirm a shell-type morphology for the SNR. In addition to the bright 
X-ray shell, we detect faint diffuse emission which extends beyond the shell 
to define the outer boundary nearly all around the SNR. Our spatially-resolved 
spectral analysis of the SNR indicates that the blast wave of G299.2$-$2.9 
is expanding into a highly inhomogeneous ISM with ambient densities varying 
over more than an order of magnitude, including several small ISM clouds 
with densities of $\sim$2$-$4 cm$^{-3}$ that have been engulfed by the 
blast wave. Bright shell emission (excluding the clouds) originates from 
shocked interstellar material with preshock densities of $n_0$ $\sim$ 
0.2$-$0.5 cm$^{-3}$. Faint extended emission beyond the shell comes from a 
lower density medium ($n_0$ $\la$ 0.1 cm$^{-3}$) also heated by the blast 
wave. These morphological and spectral characteristics suggest to us that 
the supernova exploded within a region with a large-scale gradient in ambient 
density, perhaps even near the boundary between two distinct density regions 
of the ISM.  In this scenario, the blast wave will expand at a higher velocity 
into the lower density ISM than into the higher density, which will result in 
two ``hemispheres'' with different sizes that, if viewed from the side, might 
have a morphology that is schematically shown in Fig.~\ref{fig:fig10}. 
G299.2$-$2.9 may represent a nearly face-on view of such a configuration.

How do our derived ambient density values compare with expectations? The 
density of the interstellar medium is a strong function of $z$ distance 
above the Galactic plane.  Given the SNR's location 2.9$^\circ$ below the 
plane, its $z$ distance varies from $\sim$100 pc to $\sim$500 pc for 
distances between 2 and 10 kpc. In a recent review, Ferri\'ere (2001) 
provides analytical relations for the $z$-height variation of various
constituents of the ISM in the Milky Way. Using these relations we determine 
the expected ambient number density to be $\sim$0.4 cm$^{-3}$ for $z = 100$ 
pc (corresponding to a distance of 2 kpc) and $\sim$0.03 cm$^{-3}$ for 
$z = 500$ pc (distance of 10 kpc). The midpoint distance of 5 kpc yields an 
expected ambient density of $\sim$0.1 cm$^{-3}$ that is in good agreement 
with our X-ray measurements.

The central region of G299.2$-$2.9 shows faint diffuse emission with strong 
line emission from Si, S, and Fe. This central region is most likely 
metal-rich stellar ejecta heated by the reverse shock. The ejecta are 
enriched in Si and Fe, while showing no evidence for enhancement in the 
light elements O and Ne. The inferred mass of Si in the ejecta is roughly 
half that of the Fe ejecta. This abundance pattern in the central ejecta 
suggests a Type Ia origin for the SNR. The detailed radial variation of the 
Fe and Si species, however, cannot be determined with the current data. 
Deeper X-ray observations, covering the entire SNR, would allow a more 
detailed study of this rare Galactic example of a middle-aged Type Ia 
SNR\footnote{While completing this work, we became aware of an independent 
study by L.~Pittroff et al.\ on the {\it XMM-Newton} observations of this 
same SNR. The main conclusions reached here appear consistent with their 
preliminary results.}.

We have used the X-ray temperatures, sizes, and density measurements to 
estimate the age and explosion energy of the originating SN under the 
assumption that the SNR is currently in the Sedov phase of evolution.  
This calculation was carried out separately for both the faint and bright 
portions of the shell, which provided consistent values. G299.2$-$2.9 is 
middle-aged ($\tau$ $\sim$ 4500 yr) with a supernova explosion energy of 
$E_0$ $\sim$ 1.6 $\times$ 10$^{50}$ ergs, if the SNR is located at the 
midpoint value (5 kpc) of our plausible range of distances. This value 
for the explosion energy is rather low -- most SN explosion models predict 
$10^{51}$ ergs in the kinetic energy of the ejecta. If G299.2$-$2.9 were 
as distant as 10 kpc, then the explosion energy would increase to a more 
reasonable value of $E_0$ $\sim$ 9 $\times$ 10$^{50}$ (and the age would 
become $\sim$9000 yr). However, at this distance, there would be poorer 
agreement between the X-ray--inferred ambient medium density ($n_0$ $\sim$ 
0.07 cm$^{-3}$) and the expected value of $\sim$0.03 cm$^{-3}$ (see preceding
paragraph). Given the current state of knowledge, the situation is
inconclusive. A strong constraint on the distance to G299.2$-$2.9 is going 
to be required before concluding that the originating SN was sub-energetic.

\acknowledgments

The authors thank K. Borkowski for providing us the updated NEI models.
This work was supported in part by SAO under {\it Chandra} grant SV4-74018.
POS acknowledges support from NASA contract NAS8-39073.

\clearpage

\begin{deluxetable}{lcccccc}
\footnotesize
\tablecaption{Best-fit Parameters from the Spectral Model Fit of 
the Shell Regions of G299.2$-$2.9
\label{tbl:tab1}}
\tablewidth{0pt}
\tablehead{\colhead{Region} & \colhead{$N_H$\tablenotemark{a}} & 
\colhead{$kT$\tablenotemark{a}} & \colhead{$n_et$\tablenotemark{a}} &
\colhead{$EM$\tablenotemark{a}} & \colhead{$V$\tablenotemark{b}} &
\colhead{$\chi^2$/$\nu$} \\
 & \colhead{(10$^{21}$ cm$^{-2}$)} & \colhead{(keV)} & 
\colhead{(10$^{11}$ cm$^{-3}$ s)} & \colhead{(10$^{56}$ cm$^{-3}$)} & 
\colhead{(10$^{55}$ cm$^3$)} & }
\startdata 
\vspace{1.0mm}
1 & 2.6$^{+1.1}_{-1.1}$ & 0.54$^{+0.13}_{-0.14}$ & 
1.42$^{+1.20}_{-0.42}$ & 1.03$^{+1.08}_{-0.30}$ & 45.6 & 61.3/54 \\
\vspace{1.0mm}
2 & 4.6$^{+0.7}_{-0.8}$ & 0.34$^{+0.08}_{-0.06}$ & 
3.12$^{+6.21}_{-1.24}$ & 6.09$^{+6.00}_{-3.03}$ & 10.4 & 78.7/58 \\
\vspace{1.0mm}
3 & 4.6$^{+0.7}_{-0.6}$ & 0.35$^{+0.07}_{-0.06}$ & 
5.53$^{+16.50}_{-2.15}$ & 6.84$^{+5.85}_{-2.79}$ & 0.3 & 62.2/61 \\
\vspace{1.0mm}
4 & 2.5$^{+0.6}_{-0.6}$ & 0.68$^{+0.08}_{-0.07}$ & 
0.84$^{+0.57}_{-0.30}$ & 0.70$^{+0.23}_{-0.12}$ & 24.1 & 54.0/58 \\
\vspace{1.0mm}
5 & 4.2$^{+0.6}_{-0.7}$ & 0.36$^{+0.08}_{-0.07}$ & 
$>$ 5.27 & 6.00$^{+4.35}_{-2.67}$ & 0.8 & 56.0/62 \\
\vspace{1.0mm}
6 & 3.5$^{+0.8}_{-0.7}$ & 0.56$^{+0.06}_{-0.07}$ & 
1.56$^{+0.75}_{-0.59}$ & 1.63$^{+0.83}_{-0.35}$ & 8.7 & 59.2/63 \\
\vspace{1.0mm}
7 & 3.7$^{+0.7}_{-0.8}$ & 0.40$^{+0.08}_{-0.06}$ & 
2.00$^{+1.40}_{-0.75}$ & 3.21$^{+2.42}_{-1.39}$ & 8.7 & 77.8/57 \\
\vspace{1.0mm}
8 & 3.5$^{+0.7}_{-0.5}$ & 0.53$^{+0.03}_{-0.05}$ & 
8.88$^{+3.82}_{-2.74}$ & 1.47$^{+0.46}_{-0.25}$ & 12.2 & 77.8/66 \\
\vspace{1.0mm}
9 & 3.1$^{+0.6}_{-0.6}$ & 0.71$^{+0.10}_{-0.09}$ & 
0.72$^{+0.55}_{-0.29}$ & 0.64$^{+0.28}_{-0.12}$ & 7.2 & 64.1/57 \\
\vspace{1.0mm}
10 & 3.0$^{+0.8}_{-0.6}$ & 0.52$^{+0.05}_{-0.06}$ & 
2.51$^{+1.10}_{-0.93}$ & 1.39$^{+0.62}_{-0.32}$ & 6.3 & 60.5/60 \\
\vspace{1.0mm}
11 & 3.1$^{+0.5}_{-0.4}$ & 0.58$^{+0.03}_{-0.02}$ & 
14.60$^{+7.61}_{-4.57}$ & 1.35$^{+0.20}_{-0.15}$ & 11.1 & 78.6/66 \\
\vspace{1.0mm}
12 & 1.7$^{+0.7}_{-0.5}$ & 0.56$^{+0.04}_{-0.07}$ & 
4.53$^{+1.71}_{-1.49}$ & 0.68$^{+0.23}_{-0.11}$ & 19.3 & 80.4/58 \\
\vspace{1.0mm}
13 & 2.4$^{+0.4}_{-0.4}$ & 0.64$^{+0.03}_{-0.02}$ & 
4.51$^{+1.49}_{-1.24}$ & 1.07$^{+0.14}_{-0.10}$ & 13.4 & 77.0/70 \\
\vspace{1.0mm}
14 & 2.9$^{+0.6}_{-0.5}$ & 0.57$^{+0.02}_{-0.03}$ & 
7.35$^{+2.39}_{-2.06}$ & 1.36$^{+0.25}_{-0.18}$ & 12.3 & 100.3/68 \\
\vspace{1.0mm}
15 & 3.3$^{+0.7}_{-0.6}$ & 0.57$^{+0.03}_{-0.05}$ & 
5.99$^{+3.05}_{-1.97}$ & 1.15$^{+0.33}_{-0.19}$ & 14.3 & 63.4/62 \\
\vspace{1.0mm}
16 & 3.6$^{+0.5}_{-0.5}$ & 0.63$^{+0.07}_{-0.06}$ & 
1.67$^{+0.79}_{-0.57}$ & 2.80$^{+0.77}_{-0.51}$ & 32.4 & 113.8/67 \\
17 & 3.1$^{+0.8}_{-0.6}$ & 0.88$^{+0.28}_{-0.19}$ & 
0.54$^{+0.49}_{-0.21}$ & 1.27$^{+0.68}_{-0.38}$ & 205.4 & 42.6/55 \\
\enddata

\tablenotetext{a}{The errors are estimated with the elemental abundances 
fixed at the best-fit values. The fitted mean abundances are O = 0.63$\pm$0.28, 
Ne = 0.65$\pm$0.22, Mg = 0.54$\pm$0.19, Si = 0.67$\pm$0.27, and Fe = 
0.66$\pm$0.28.}
\tablenotetext{b}{Emission volume (assuming $d$ = 5 kpc) used to calculate 
the density for each region.}
\end{deluxetable}

\begin{deluxetable}{cccccccc}
\footnotesize
\tablecaption{Best-fit Parameters from the Spectral Model Fit of 
the Central Nebula of G299.2$-$2.9
\label{tbl:tab2}}
\tablewidth{0pt}
\tablehead{\colhead{$N_H$\tablenotemark{a}} & 
\colhead{$kT$\tablenotemark{a}} & \colhead{$n_et$\tablenotemark{a}} &
\colhead{$EM$\tablenotemark{a}} & \colhead{Si\tablenotemark{b}} &
\colhead{S\tablenotemark{b}} & \colhead{Fe\tablenotemark{b}} &
\colhead{$\chi^2$/$\nu$} \\
\colhead{(10$^{21}$ cm$^{-2}$)} & \colhead{(keV)} & 
\colhead{(10$^{11}$ cm$^{-3}$ s)} & \colhead{(10$^{56}$ cm$^{-3}$)} &
& & & } 
\startdata 
\vspace{1.0mm}
3.7$^{+0.3}_{-0.3}$ & 1.36$^{+0.03}_{-0.13}$ & 0.13$^{+0.03}_{-0.03}$ & 
0.25$^{+0.08}_{-0.06}$ & 7.25$^{+1.84}_{-1.42}$ & 19.59$^{+6.69}_{-5.31}$ 
& 7.72$^{+2.32}_{-1.79}$ & 85.6/60 \\
\enddata

\tablenotetext{a}{The errors are estimated with 
the elemental abundances fixed at the best-fit values.}
\tablenotetext{b}{The best-fit abundances with respect to the solar.
The fitted abundances for other elements (O, Ne, and Mg) are low ($\la$0.6)
and have been fixed before estimating the 1$\sigma$ uncertainties for
Si, S, and Fe.}

\end{deluxetable}

\begin{deluxetable}{lccc}
\footnotesize
\tablecaption{SNR Parameters of G299.2$-$2.9
\label{tbl:tab3}}
\tablewidth{0pt}
\tablehead{\colhead{Parameter} & \colhead{Bright Shell} & 
\colhead{Faint Rim} & \colhead{Entire SNR\tablenotemark{a}}}
\startdata 
\vspace{1.0mm}
$kT$ (keV) & 0.52$\pm$0.04\tablenotemark{b} & 0.67$\pm$0.04\tablenotemark{c} & - \\
$R$ ($d_5$ pc) & 7.2 & 9.2 & - \\
\hline

$n_{\rm e}$ ($f^{-{1\over2}}$ $d^{-{1\over2}}_5$ cm$^{-3}$) & 1.58$\pm$0.16 
& 0.50$\pm$0.06 & - \\
$n_0$ ($f^{-{1\over2}}$ $d^{-{1\over2}}_5$ cm$^{-3}$) & 0.33$\pm$0.03 & 0.11$\pm$0.01 
& - \\
$E_0$ ($d^{{5\over2}}_5$$f^{-{1\over2}}$ 10$^{50}$ ergs) & 1.71$\pm$0.22 & 
1.55$\pm$0.20 & 1.63$\pm$0.21 \\
Age ($d_5$ yr) & 4200$\pm$170 & 4700$\pm$140 & 4500$\pm$160 \\
$M_{\rm swept}$ ($f^{1\over2}$ $d^{{5\over2}}_5$ $M_{\odot}$) & 12.8$\pm$1.3 & 
8.8$\pm$1.0 & 10.8$\pm$1.1 \\
\enddata

\tablecomments{1$\sigma$ uncertainties are quoted.}
\tablenotetext{a}{Mean values of the two assumed cases.}
\tablenotetext{b}{On average based on the spectral fits from the bright
shell (i.e., regions 2, 6$-$7, 10, 13$-$14, and 16). The brightest northeastern
knots (regions 3 and 5) are excluded.}
\tablenotetext{c}{On average based on the spectral fits from the faint outer
boundary (i.e., regions 1, 4, 12, and 17).}

\end{deluxetable}

\end{document}